\documentclass{llncs}
\pdfoutput=1

\usepackage[pdftex]{graphicx}
\usepackage[english]{babel}
\usepackage{amsmath}
\usepackage{color}
\usepackage{amsfonts}
\usepackage{acronym}
\usepackage{array}
\usepackage{fixltx2e}
\usepackage{acronym}
\usepackage{listings}
\usepackage{url}
\usepackage[nocompress]{cite}
\usepackage{amssymb}
\usepackage{diagbox}
\usepackage{array}
\usepackage{subfig}
\usepackage{adjustbox}
\usepackage{pgf}
\usepackage{tikz}
\usepackage{courier}

\lstdefinelanguage{Modelica}
{
  morekeywords = {model,discrete,when,parameter,initial,equation,der,then,end,reinit,pre},
  sensitive = true,
  escapeinside={/+}{+/},
}
\lstnewenvironment{Modelica}
{\lstset{language={Modelica}
}} {}

\begin{document}

\title{A Brief Overview of the KTA WCET Tool}
\titlerunning{}
\author{David Broman}
\authorrunning{David Broman} 
\institute{KTH Royal Institute of Technology\\Sweden\\
\email{dbro@kth.se}}

\maketitle

\begin{abstract}
KTA (KTH's timing analyzer) is a research tool for performing timing
analysis of program code. The currently available toolchain can
perform two different kinds of analyses: i) exhaustive fine-grained
timing analysis, where timing information can be provided between
arbitrary timing program points within a function, and ii) abstract
search-based timing analysis, where the tool can perform
optimal worst-case execution time (WCET) analysis. The latter is based
on a technique that combines divide-and-conquer search and abstract
interpretation. The tool is under development and currently supports a
subset of the MIPS instruction set architecture.
\end{abstract}


\section{Introduction}
\label{sec:intro}

The \emph{worst-case execution time (WCET)}
problem~\cite{WilhelmEtAl:2008,PuschnerBurns:2000} is an important
research area within the context of real-time systems. There exist
many tools and techniques for static WCET
analyis~\cite{GustafssonEtAl:2006,ColinPuaut:2000,FerdinandWilhelm:1999,LiLiangMitra:2007,LiMalik:1997,Ballabriga:2010},
for measurement-based and probabilistic
approaches~\cite{BernatEtAl:2003,DavidPuaut:2004}, and alternative
approaches that are based on simplified
hardware~\cite{PuautPais:2007,KimEtAl:2014,ZimmerEtAl:2014,RochangeEtAl:2014,AxerEtAl:2014,SchoeberlEtAl:2010,KimEtAl:2017}. Recent
work also targets the challenging problem of multicore WCET analysis
~\cite{ChattopadhyayEtAl:2014,LiMitraEtAl:2009,TanMooney:2007,KastnerEtAl:2012,PellizzoniEtAl:2010,MancusoEtAl:2015}. Although
several of the state-of-the-art tools can estimate a safe WCET bound
at the function level, there is currently no existing tool that can
provide guaranteed optimal WCET values between specific program
points. The aim of the KTA tool is to provide such optimal and
guaranteed fine-grained analysis. The toolchain is available as open source\footnote{\url{https://github.com/timed-c/kta}}.

This short paper gives a brief overview of the key ideas and history
behind the KTA tool. Section~\ref{sec:overview} gives an overview of
the main use cases and objectives. Section~\ref{sec:architecture}
describes the main architectural components, and
Section~\ref{sec:futurework} discusses some future research
directions.

\section{Background and Objectives}
\label{sec:overview}

The early work of the toolchain started in 2013 during the time when
the author of this paper worked at UC Berkeley within the PRET
project~\cite{EdwardsLee:2007,LiuEtAl:2012,ZimmerEtAl:2014}. As part
of the vision of a toolchain~\cite{BromanEtAlPretInf:2013}, the
objective was to support WCET analysis for the RISC-V instruction set
architecture (ISA) and to perform parts of the analysis within the
LLVM~\cite{LattnerAdve:2004} toolchain. However, when the author
moved to KTH, the focus shifted to low-level analysis at the machine
code level.  For this purpose, the MIPS ISA was used
instead, partially because of the need  to support the analysis of 
off-the-shelf hardware. The MIPS architecture was also chosen because
of its rather simple structure and its common use in education.

Today, there are two separate timing analysis
methods, with the following separate objectives.

\begin{enumerate}
\item \textbf{Exhaustive fine-grained timing analysis.}  The objective
  of this fine-grain\-ed analysis is to enable both WCET analysis and
  best-case execution time (BCET) analysis between arbitrary program
  points within a function. The current version of the work is
  primarily used in the context of interactive timing
  analysis~\cite{FuhrmannEtAl:2016}, where a graphical modeling tool
  can be used to identify hotspots of the model that are contributing
  significantly to the WCET path. The current version of the
  fine-grained analysis is based on exhaustively searching all paths
  between programming points.  As a consequence, this approach is not
  scalable, but it has been very useful for identifying the
  fine-grained analysis methodology.  We see it as future work to
  combine this fine-grained timing-point methodology with the next
  approach that is based on abstract
  interpretation~\cite{CousotCousot:1977,Cousot:2001}.
  
\item \textbf{Abstract search-based timing analysis.} The objective of
  the abstract search-based WCET analysis is to perform highly
  scalable WCET analysis that returns optimal WCET values. By optimal
  we mean WCET estimates that are sound and equal to the actual
  WCET. Note that the estimated WCET value is only optimal with
  respect to the \emph{model} of the hardware platform, and not
  necessarily with respect to the hardware itself. That is, we must assume that the
  model of the hardware is sound, but this is typically hard to
  actually prove in practice.  The key aspects of this analysis are i)
  the analysis is performed using a technique based on abstract
  interpretation, ii) it performs a combined-phase analysis, where
  program-flow analysis, microarchitecture analysis, and global-bound
  analysis are combined into one global phase, and iii) the optimal
  WCET value is computed using an abstract search-based method, which
  is based on a divide-and-conquer approach.
\end{enumerate}

\noindent Between the years 2013 and 2016, the work on the KTA tool
was performed by David Broman. In the year 2017, the Master's student
Rodothea-Myrsini Tsoupidi started to extend the KTA tool with the
support for cache analysis and pipeline analysis. The design and
implementation described in this paper only reflects the work done by
David prior and during 2017. The microarchitecture extensions
developed in Tsoupidi's Master's thesis are briefly mentioned as future
work in Section~\ref{sec:futurework}.

\begin{figure*}[!b]
\center
\includegraphics[width=1.0\textwidth]{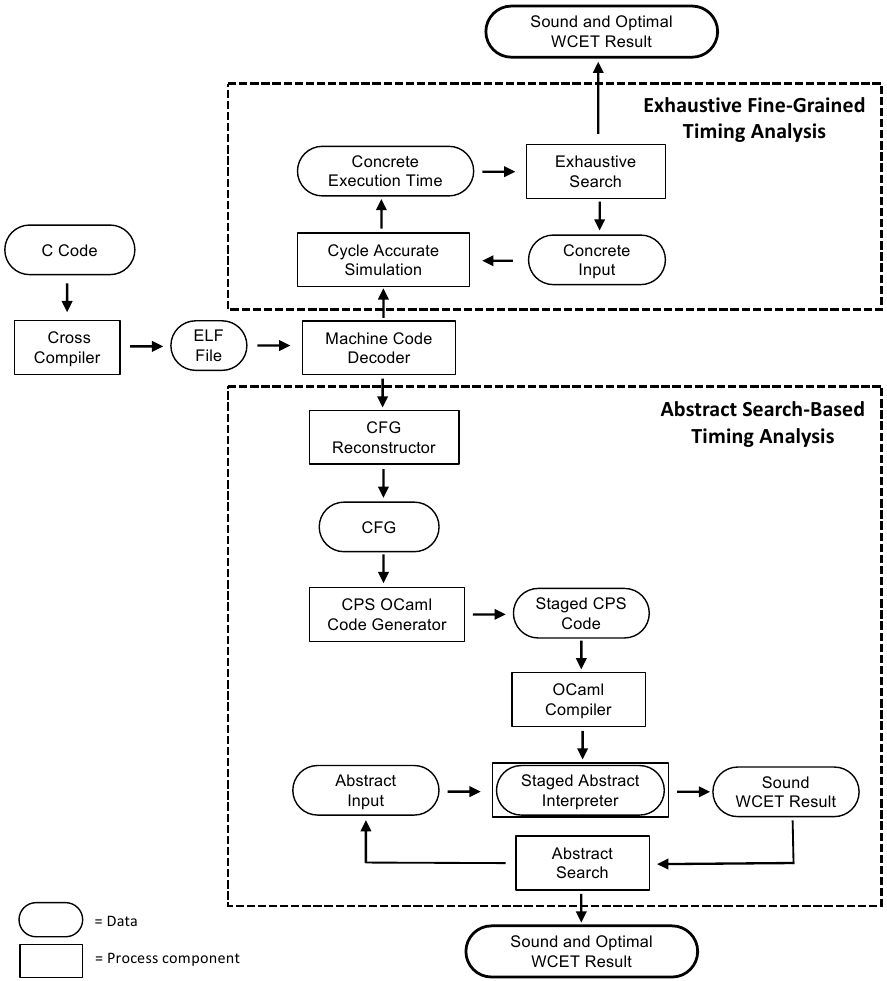}
\caption{An architectural overview of the KTA tool.}
\label{fig:arch}
\end{figure*}

\section{Architecture Overview}
\label{sec:architecture}

Fig.~\ref{fig:arch} depicts the main components and flow of
information within the KTH tool. The picture shows the two main
analysis flows: i) \emph{exhaustive fine-grained timing analysis} (top
part of the figure), and ii) \emph{abstract search-based timing
  analysis} (bottom part). The boxes represent processing components,
and the rounded boxes represent \emph{data}. The main input to the
tool is a C program (left part of the figure), together with a set of
parameters (not shown in the figure). The C code is first compiled
using a standard \emph{cross compiler} for the C programming
language. In our case, we used a \verb|gcc| variant that targets the
MIPS instruction set architecture. The cross compiler generates an ELF
(executable and linkable format) file. The different sections
(\verb|.text|, \verb|.data|, etc.) of the ELF file are decoded. In
particular, the MIPS machine code is decoded into an internal
format. All code is written in OCaml and compiled using the OCaml compiler
version 4.05.0.

The rest of this section describes the main ideas of the
two different analyses.

\subsection{Exhaustive Fine-Grained Timing Analysis}
The exhaustive fine-grained timing analysis (top part of
Fig.~\ref{fig:arch}) takes as input the decoded machine code and
returns a sound and optimal WCET result, if the search terminates
within a specific time limit.

The exhaustive search consists conceptually of a loop where a
\emph{cycle accurate simulation} is first performed with some selected
\emph{concrete input}. The output from the simulation is a
\emph{concrete execution time} value, that is then used by the
\emph{exhaustive search} component to select the next concrete
input. This procedure continues until all program inputs have been
explored. The procedure stores timing information at predefined timing
points, which are then later used for computing the WCET and BCET
values between timing points. Please see the paper by Fuhrmann
\emph{et al.}\cite{FuhrmannEtAl:2016} for more information.

\subsection{Abstract Search-Based Timing Analysis}

The main flow of the abstract search-based timing analysis is shown in
the bottom part of Fig.~\ref{fig:arch}. In the first step, the control
flow graph (CFG) of the machine code is reconstructed. The CFG is then
the input to a \emph{CPS OCaml code generator} that outputs OCaml code in
continuous-passing style (CPS). This staged CPS machine code is
then compiled (using an \emph{OCaml compiler}) into a \emph{staged
  abstract interpreter}. This is one of the key ideas of KTH: the
machine code of the program that is going to be analyzed is in fact
translated into a program that performs abstract interpretation by
executing the machine code abstractly.

Note that the data item \emph{staged abstract interpreter} is depicted
both as a normal box (a process component) and a rounded box (a data
item). This means that the staged interpreter is compiled into a
binary artifact that is then executed in the circular  
control-flow graph (shown at the bottom of the figure).

The abstract-search phase is performed as follows. First, the component called
\emph{abstract search} selects some abstract input. By
abstract input we mean a set of values (typically an interval) that
represents a subset of the input space. The \emph{staged abstract
  interpreter} performs an analysis phase based on this input, and
generates a sound (but not necessarily optimal) WCET
result\footnote{Note that the tool can potentially generate BCET
  values as well, but it is not completely implemented in the current
  version.}. The WCET result is then used as input again to the
  abstract search component, that selects another relevant
  \emph{abstract input}. The abstract search algorithm performs a
  divide-and-conquer analysis to enable faster search of the optimal
  WCET value.

  Note that the abstract search is \emph{bounded}, which means that
  the abstract interpretation terminates if a simulated max-time value is
  reached. This is actually natural in a real-time scheduling setting
  because we can often assume that the maximal reasonable WCET value
  is the period of a task. Hence, we get a termination definition that
  is not directly dependent on the analysis time.

\section{Future Research}
\label{sec:futurework}

As stated before, this KTA tool can so far be seen as work in
progress. However, we are currently extending the tool in a number of
aspects. More specifically, the following can be seen as prioritized
ongoing and future work:
\begin{itemize}
\item The tool is currently being extended to include more complicated
  micro architectures. In particular, Tsoupidi's ongoing Master's thesis is
  focusing on extending the KTA tool with sound cache analysis and
  sound pipeline analysis.
\item We would like to inspect if the tool can be extended with the
  non-relational Polyhedra domain~\cite{SinghEtAl:2017}.
\item We will investigate how the tool can be extended to also support
  multicore analysis.
\item An interesting problem would be to combine the fine-grained
  timing analysis, with the above presented abstract-search based
  method.
\end{itemize}

\section{Conclusions}
\label{sec:conclusion}
In this paper, we give a brief overview of the KTA tool. In
particular, the paper describes two main approaches of timing analysis
that are available in KTA: i) exhaustive fine-grained timing analysis,
and ii) abstract search-based timing analysis. We content that the
latter approach---where all phases in traditional WCET
analysis are combined into one pass---can be a serious alternative
approach to traditional WCET analysis.

\section*{Acknowledgments}

This project is financially supported by the Swedish Foundation for
Strategic Research (FFL15-0032).  The research work has previously
been funded by the Swedish Research Council (\#623-2011-955 and
\#623-2013-8591). I would like to thank Rodothea-Myrsini Tsoupidi and
Saranya Natarajan for comments on the final version of this paper.

\newpage
\bibliographystyle{plain}
\bibliography{references}

\begin{thebibliography}{10}

\bibitem{AxerEtAl:2014}
Philip Axer, Rolf Ernst, Heiko Falk, Alain Girault, Daniel Grund, Nan Guan,
  Bengt Jonsson, Peter Marwedel, Jan Reineke, Christine Rochange, Maurice
  Sebastian, Reinhard~Von Hanxleden, Reinhard Wilhelm, and Wang Yi.
\newblock {Building Timing Predictable Embedded Systems}.
\newblock {\em ACM Transactions on Embedded Computing Systems (TECS)},
  13(4):82:1--82:37, March 2014.

\bibitem{Ballabriga:2010}
Cl{\'e}ment Ballabriga, Hugues Cass{\'e}, Christine Rochange, and Pascal
  Sainrat.
\newblock {OTAWA: An open toolbox for adaptive WCET analysis}.
\newblock In {\em Software Technologies for Embedded and Ubiquitous Systems},
  volume 6399 of {\em LNCS}, pages 35--46. Springer, 2010.

\bibitem{BernatEtAl:2003}
Guillem Bernat, Antoine Colin, and Stefan Petters.
\newblock {pWCET: A tool for probabilistic worst-case execution time analysis
  of real-time systems}.
\newblock In {\em Proceedings of the 3rd International Workshop on Worst-Case
  Execution Time Analysis (WCET)}, pages 21--38, 2003.

\bibitem{BromanEtAlPretInf:2013}
David Broman, Michael Zimmer, Yooseong Kim, Hokeun Kim, Jian Cai, Aviral
  Shrivastava, Stephen~A. Edwards, and Edward~A. Lee.
\newblock {Precision Timed Infrastructure: Design Challenges}.
\newblock In {\em Proceedings of the Electronic System Level Synthesis
  Conference (ESLsyn)}. IEEE, 2013.

\bibitem{ChattopadhyayEtAl:2014}
Sudipta Chattopadhyay, Lee~Kee Chong, Abhik Roychoudhury, Timon Kelter, Peter
  Marwedel, and Heiko Falk.
\newblock A unified wcet analysis framework for multicore platforms.
\newblock {\em ACM Transactions on Embedded Computing Systems (TECS)},
  13(4s):124, 2014.

\bibitem{ColinPuaut:2000}
Antoine Colin and Isabelle Puaut.
\newblock {Worst case execution time analysis for a processor with branch
  prediction}.
\newblock {\em Real-Time Systems}, 18(2-3):249--274, 2000.

\bibitem{Cousot:2001}
Patrick Cousot.
\newblock {Abstract interpretation based formal methods and future challenges}.
\newblock In {\em Informatics}, volume 2000 of {\em LNCS}, pages 138--156.
  Springer, 2001.

\bibitem{CousotCousot:1977}
Patrick Cousot and Radhia Cousot.
\newblock Abstract interpretation: a unified lattice model for static analysis
  of programs by construction or approximation of fixpoints.
\newblock In {\em Proceedings of the 4th ACM SIGACT-SIGPLAN symposium on
  Principles of programming languages (POPL)}, pages 238--252, New York, USA,
  1977. ACM Press.

\bibitem{DavidPuaut:2004}
Laurent David and Isabelle Puaut.
\newblock Static determination of probabilistic execution times.
\newblock In {\em Proceedings of the 16th Euromicro Conference on Real-Time
  Systems (ECRTS)}, pages 223--230. IEEE, 2004.

\bibitem{EdwardsLee:2007}
Stephen~A. Edwards and Edward~A. Lee.
\newblock The case for the precision timed (pret) machine.
\newblock In {\em Proceedings of the 44th annual conference on Design
  automation}, pages 264 -- 265, June 2007.

\bibitem{FerdinandWilhelm:1999}
Christian Ferdinand and Reinhard Wilhelm.
\newblock {Efficient and precise cache behavior prediction for real-time
  systems}.
\newblock {\em {Real-Time Systems}}, 17(2):131--181, 1999.

\bibitem{FuhrmannEtAl:2016}
Insa Fuhrmann, David Broman, Reinhard~Von Hanxleden, and Alexander
  Schulz-Rosengarten.
\newblock {Time for Reactive System Modeling: Interactiave Timing Analysis with
  Hotspot Highlighting}.
\newblock In {\em Proceedings of the 24th International Conference on Real-Time
  Networks and Systems (RTNS 2016)}. ACM, 2016.

\bibitem{GustafssonEtAl:2006}
Jan Gustafsson, Andreas Ermedahl, Christer Sandberg, and Bj\"orn Lisper.
\newblock {Automatic derivation of loop bounds and infeasible paths for WCET
  analysis using abstract execution}.
\newblock In {\em Proceedings of the 27th IEEE International Real-Time Systems
  Symposium (RTSS)}, pages 57--66. IEEE, 2006.

\bibitem{KastnerEtAl:2012}
Daniel K{\"a}stner, Marc Schlickling, Markus Pister, Christoph Cullmann, Gernot
  Gebhard, Reinhold Heckmann, and Christian Ferdinand.
\newblock Meeting real-time requirements with multi-core processors.
\newblock In {\em Proceedings of the Internationcal Conference on Computer
  Safety, Reliability, and Security (SAFECOMP)}, volume 7613 of {\em LNCS},
  pages 117--131. Springer, 2012.

\bibitem{KimEtAl:2014}
Yooseong Kim, David Broman, Jian Cai, and Aviral Shrivastaval.
\newblock {WCET-Aware Dynamic Code Management on Scratchpads for
  Software-Managed Multicores}.
\newblock In {\em Proceedings of the 20th IEEE Real-Time and Embedded
  Technology and Application Symposium (RTAS)}. IEEE, 2014.

\bibitem{KimEtAl:2017}
Yooseong Kim, David Broman, and Aviral Shrivastava.
\newblock {WCET-Aware Function-Level Dynamic Code Management on Scratchpad
  Memory}.
\newblock {\em {ACM Transactions on Embedded Computing Systems}},
  16(4):112:1--112:26, May 2017.

\bibitem{LattnerAdve:2004}
Chris Lattner and Vikram Adve.
\newblock {LLVM: A Compilation Framework for Lifelong Program Analysis \&
  Transformation}.
\newblock In {\em Proceedings of the International Symposium on Code Generation
  and Optimization (CGO'04)}. IEEE, 2004.

\bibitem{LiLiangMitra:2007}
Xianfeng Li, Yun Liang, Tulika Mitra, and Abhik Roychoudhury.
\newblock {Chronos: A timing analyzer for embedded software}.
\newblock {\em {Science of Computer Programming}}, 69(1):56--67, 2007.

\bibitem{LiMalik:1997}
Y-TS Li and Sharad Malik.
\newblock Performance analysis of embedded software using implicit path
  enumeration.
\newblock {\em {IEEE Transactions on Computer-Aided Design of Integrated
  Circuits and Systems}}, 16(12):1477--1487, 1997.

\bibitem{LiMitraEtAl:2009}
Yan Li, Vivy Suhendra, Yun Liang, Tulika Mitra, and Abhik Roychoudhury.
\newblock {Timing analysis of concurrent programs running on shared cache
  multi-cores}.
\newblock In {\em Proceedings of the Real-Time Systems Symposium (RTSS)}, pages
  57--67. IEEE, 2009.

\bibitem{LiuEtAl:2012}
Isaac Liu, Jan Reineke, David Broman, Michael Zimmer, and Edward~A. Lee.
\newblock {A PRET Microarchitecture Implementation with Repeatable Timing and
  Competitive Performance}.
\newblock In {\em Proceedings of the 30th IEEE International Conference on
  Computer Design (ICCD 2012)}, pages 87--93. IEEE, 2012.

\bibitem{MancusoEtAl:2015}
Renato Mancuso, Rodolfo Pellizzoni, Marco Caccamo, Lui Sha, and Heechul Yun.
\newblock Wcet (m) estimation in multi-core systems using single core
  equivalence.
\newblock In {\em Proceedings of the 27th Euromicro Conference on Real-Time
  Systems (ECRTS)}, pages 174--183. IEEE, 2015.

\bibitem{PellizzoniEtAl:2010}
Rodolfo Pellizzoni, Andreas Schranzhofer, Jian-Jia Chen, Marco Caccamo, and
  Lothar Thiele.
\newblock Worst case delay analysis for memory interference in multicore
  systems.
\newblock In {\em Proceedings of the Conference on Design, Automation and Test
  in Europe (DATE)}, pages 741--746, 2010.

\bibitem{PuautPais:2007}
Isabelle Puau and Christophe Pais.
\newblock Scratchpad memories vs locked caches in hard real-time systems: a
  quantitative comparison.
\newblock In {\em Proceedings of the Design, Automation \& Test in Europe
  Conference \& Exhibition (DATE)}, pages 1--6. IEEE, 2007.

\bibitem{PuschnerBurns:2000}
Peter Puschner and Alan Burns.
\newblock Guest editorial: A review of worst-case execution-time analysis.
\newblock {\em Real-Time Systems}, 18(2):115--128, 2000.

\bibitem{RochangeEtAl:2014}
Christine Rochange, Pascal Sainrat, and Sascha Uhrig.
\newblock {\em Time-Predictable Architectures}.
\newblock John Wiley \& Sons, 2014.

\bibitem{SchoeberlEtAl:2010}
Martin Schoeberl, Wolfgang Puffitsch, Rasmus~Ulslev Pedersen, and Benedikt
  Huber.
\newblock {Worst-case execution time analysis for a Java processor}.
\newblock {\em {Software: Practice and Experience}}, 40(6):507--542, 2010.

\bibitem{SinghEtAl:2017}
Gagandeep Singh, Markus P\"{u}schel, and Martin Vechev.
\newblock {Fast Polyhedra Abstract Domain}.
\newblock In {\em {Proceedings of the 44th ACM SIGPLAN Symposium on Principles
  of Programming Languages}}, POPL 2017, pages 46--59. ACM, 2017.

\bibitem{TanMooney:2007}
Yudong Tan and Vincent Mooney.
\newblock Timing analysis for preemptive multitasking real-time systems with
  caches.
\newblock {\em ACM Transactions on Embedded Computing System}, 6(1), February
  2007.

\bibitem{WilhelmEtAl:2008}
Reinhard Wilhelm, Jakob Engblom, Andreas Ermedahl, Niklas Holsti, Stephan
  Thesing, David Whalley, Guillem Bernat, Christian Ferdinand, Reinhold
  Heckmann, Tulika Mitra, Frank Mueller, Isabelle Puaut, Peter Puschner, Jan
  Staschulat, and Per Stenstr{\"o}m.
\newblock {The Worst-Case Execution-Time Problem - Overview of Methods and
  Survey of Tools}.
\newblock {\em ACM Transactions on Embedded Computing Systems}, 7:36:1--36:53,
  May 2008.

\bibitem{ZimmerEtAl:2014}
Michael Zimmer, David Broman, Chris Shaver, and Edward~A. Lee.
\newblock {FlexPRET: A Processor Platform for Mixed-Criticality Systems}.
\newblock In {\em Proceedings of the 20th IEEE Real-Time and Embedded
  Technology and Application Symposium (RTAS)}, pages 101--110. IEEE, 2014.

\end{thebibliography}

\end{document}